\documentclass[letterpaper, 10 pt, conference]{IEEEtran}

\usepackage{tabularx}
\usepackage[noend]{algpseudocode}

\usepackage{algorithm}
\usepackage{algorithmicx}

\usepackage{subfigure}
\usepackage{lipsum} 
\usepackage{arydshln}
\usepackage[dvips]{color}
\usepackage{comment}
\usepackage{todonotes}
\usepackage{epsf}
\usepackage{epsfig}
\usepackage{times}
\usepackage{epsfig}
\usepackage{graphicx}
\usepackage{bbold}
\usepackage{mathtools}
\usepackage{mathrsfs}
\usepackage{amssymb}
\usepackage{pdfpages}
\usepackage{epstopdf}

\usepackage{amsmath,amsfonts,amssymb,amsthm}
\usepackage{mathtools}
\usepackage{commath}
\usepackage[sc,osf]{mathpazo}

\usepackage{soul} 

\newcommand{\mtx}[1]{\mathbf{#1}}

\usepackage{amsmath}
\usepackage{dsfont}
\usepackage{lettrine} 

\usepackage{amsmath,epsfig,amssymb,algpseudocode,amsthm,cite,url}
\usepackage{caption}
\usepackage{tikz}
\usetikzlibrary{shapes,positioning,calc}
\usepackage{tikz-qtree}
\usepackage{tikzscale}
\usepackage{siunitx}
\usepackage{booktabs}

\allowdisplaybreaks
\usepackage{csquotes}
\usepackage{verbatim}
\usepackage[english]{babel}
\usepackage{amsmath,amssymb}

\DeclareMathOperator*{\argmax}{arg\,max}

\usepackage{multirow}
\usepackage{rotating}
\usepackage{glossaries}

\usepackage[justification=RaggedRight]{caption}
\usepackage{verbatim}

\begin{document}
	%
	 \title{Learning Robust Scheduling \\ with Search and Attention\vspace{-0.3cm}}
	\IEEEoverridecommandlockouts
	\author{\IEEEauthorblockN{David Sandberg, Tor Kvernvik, Francesco Davide Calabrese\\}
			Ericsson, Stockholm, Sweden\\
           Email: \{david.sandberg, tor.kvernvik, francesco.davide.calabrese\}@ericsson.com

\vspace{-0.2cm}

	}
	\maketitle
	
\IEEEpeerreviewmaketitle

\newglossaryentry{lte}{name=LTE, description={Long-Term Evolution}}
\newglossaryentry{nr}{name=NR, description={New Radio}}
\newglossaryentry{tti}{name=TTI, description={Transmit Time Interval}}

\vspace{-1cm}	
\begin{abstract}
Allocating physical layer resources to users based on channel quality, buffer size, requirements and constraints represents one of the central optimization problems in the management of radio resources.
The solution space grows combinatorially with the cardinality of each dimension making it hard to find optimal solutions using an exhaustive search or even classical optimization algorithms given the stringent time requirements.
This problem is even more pronounced in MU-MIMO scheduling where the scheduler can assign multiple users to the same time-frequency physical resources.
Traditional approaches thus resort to designing heuristics that trade optimality in favor of feasibility of execution.
In this work we treat the MU-MIMO scheduling problem as a tree-structured combinatorial problem and, borrowing from the recent successes of AlphaGo Zero, we investigate the feasibility of searching for the best performing solutions using a combination of Monte Carlo Tree Search and Reinforcement Learning.
To cater to the nature of the problem at hand, like the lack of an intrinsic ordering of the users as well as the importance of dependencies between combinations of users, we make fundamental modifications to the neural network architecture by introducing the self-attention mechanism. We then demonstrate that the resulting approach is not only feasible but vastly outperforms state-of-the-art heuristic-based scheduling approaches in the presence of measurement uncertainties and finite buffers.

\end{abstract}
\vspace{-0.1cm}
\section{Introduction}
\vspace{-0.1cm}

Multi-User MIMO (MU-MIMO) is a technology that has the potential to considerably
improve the spectral efficiency of wireless systems by superimposing different users transmissions over the same time/frequency resources.
However, this comes at the cost of reduced Signal to Interference and Noise Ratio (SINR) for the co-scheduled users caused by interference between the users that can severely limit
the achievable gains unless adequate signal processing and transmission techniques are used to mitigate it.
Resource allocation strategies, in deciding which users to schedule together, play a fundamental role in maximizing the gain of non-orthogonal access techniques like MU-MIMO.
However, a variety of variables need to be taken into account when performing such allocation: buffer state, QoS requirements, signaling limitations, power limitations, channel conditions, etc.
In particular, the channel, whose characteristics vary greatly in time, frequency and space, represents a fundamental variable to take into account if considerable gains should be achieved.
For example, users with similar channels are more likely to interfere with each other and should therefore not be multiplexed in the time/frequency domain.
At the same time the variability of the channel, together with the need to consider other variables as well as numerous constraints
dramatically increase the complexity of the solution. Therefore, the common approach in the literature has been to design suboptimal heuristic algorithms that balance complexity and performance.

In \cite{MumimoResourceAlloc}, the authors provide an excellent in-depth review of prior work on scheduling for MU-MIMO. Here, current scheduling methods are split into \textit{Classical Optimization}, \textit{Metaheuristic Optimization} and \textit{Aggregated Utility Based Selection}. The first two classes suffer from high computational complexity and are considered too demanding for real-time operation. The third class contains the \textit{Direct} methods that directly maximize a utility function, as well as the \textit{Indirect} methods that first calculate a compatibility metric and then use it for optimization. This last class is where most real-time implementations, including the baseline in this work, belong.

In this work we treat scheduling as a tree-structured combinatorial problem.
Imposing a tree structure on the scheduling problem opens the door for a variety of well established tree-search techniques that are often combined with domain-specific heuristics to speed up the search.
In our work we take an approach based on Monte Carlo Tree Search (MCTS) and model-based Reinforcement Learning (RL) that enables us to break free of heuristics and, instead, "learn to search" for solutions within the tree. This approach is heavily inspired by \mbox{AlphaGo} Zero \cite{Silver2017MasteringTG}, inarguably regarded as one of the most significant achievements in AI in recent years. However, it is significantly modified to cater to the nature of the problem at hand. In particular, the lack of an intrinsic ordering of users and the importance of dependencies between combinations of users calls for adaptations in the neural network architecture. By introducing self-attention in the neural network we design an architecture that can handle the combinatorial explosion of user permutations efficiently.
The result is an effective scheduler that vastly outperforms advanced heuristic-based scheduling algorithms in the presence of measurement uncertainties and finite buffers.


\vspace{-0.15cm}
\section{Scheduling Framework}\label{scheduling_framework}
\vspace{-0.1cm}

In the following we formulate the scheduling process as a Markov Decision Process (MDP) applied to a tree. Then we show how to use MCTS combined with RL to solve such MDP.

\vspace{-0.1cm}
\subsection{Scheduling as a Markov Decision Process}\label{mdp}
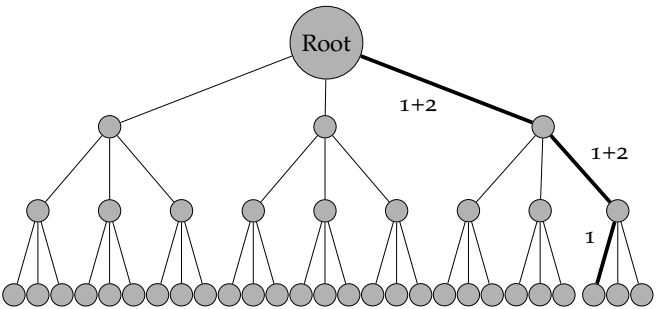
\begin{figure}[!ht]
\vspace{-0.2cm}
\centering
\resizebox{3.45in}{!}{\begin{tikzpicture}[every tree node/.style={draw,circle, minimum size=0.3em,fill=black!30},
   level distance=1.25cm,sibling distance=.01cm, 
   edge from parent path={(\tikzparentnode) -- (\tikzchildnode)}]
\Tree[.\node (Root) {Root};
    [.{} 
      [.{} [.{} ] [.{} ] [.{} ]]
      [.{} [.{} ] [.{} ] [.{} ]]
      [.{} [.{} ] [.{} ] [.{} ]]
    ]
    [.{} 
      [.{} [.{} ] [.{} ] [.{} ]]
      [.{} [.{} ] [.{} ] [.{} ]]
      [.{} [.{} ] [.{} ] [.{} ]]
    ]
    \edge [ultra thick] node[auto=right] {1+2};
    [.{} 
      [.{} [.{} ] [.{} ] [.{} ]]
      [.{} [.{} ] [.{} ] [.{} ]]
      \edge [ultra thick] node[auto=left] {1+2};
      [.{} \edge [ultra thick] node[auto=right] {1};  [.{} ] [.{} ] [.{} ]]
    ]
    ]
\end{tikzpicture}}
\caption{The search starts from the root state with no subband allocated. In the example the first and second subbands are allocated to both user 1 and 2, whereas the third subband is allocated to user 1 alone.}
\vspace{-0.2cm}
\label{fig:search_tree}
\end{figure}

An episodic MDP can be summarized as
\begin{equation*}
s',r\gets E(s,a)
\end{equation*}
where $E$ is an episodic simulator that given the current state~$s$ and the action taken in that state~$a$ as input, returns a new state~$s'$ and a corresponding reward~$r$.
The sequential formulation can be easily imposed within one scheduling episode by iterating over the subbands following an arbitrary order and selecting the users to assign to each subband.
The mapping of users to subbands can be visualized as a tree (see Fig.~\ref{fig:search_tree}) with the depth equal to the number of subbands $N_\text{subband}$. 
The branching factor of the tree corresponds to the cardinality of the action space $\mathcal{A}$, that is, all the possible ways of selecting a maximum of $M$ users out of the total $N_\text{user}$, can be computed as:
\begin{equation*}
|\mathcal{A}| = {N_\text{user}+1\choose M}
\end{equation*}

The number of tree leaves (i.e., the terminal states of each episode, where the reward is collected) represents the cardinality of the solution space $\mathcal{S}$, computed as
\begin{equation*}
|\mathcal{S}|=|\mathcal{A}|^{N_\text{subband}}
\end{equation*}
For a scenario with 4 users and 10 subbands the size of the solution space is $|\mathcal{S}|=10^{10}$, well beyond what can be handled by any traditional search technique.

The state consists of two parts. The first part is an environment state which contains channel state (for each combination of user and subband) and buffer state (per user). This part is sampled when the environment is created and does not change during the episode. The second part is the episode state which is more dynamic and changes depending on the actions taken earlier in the episode. This part consists of an allocation matrix that indicates which users have been assigned to each subband. Both the environment state and the episode state are required for a policy to take correct actions.

It is important to note that even if the search is done sequentially over subbands, the state contains the information needed to make globally optimal decisions.

\vspace{-0.1cm}
\subsection{Sampling of the Environment State}\label{state_sampling}
The environment state, consisting of the channel and the buffer state, is sampled when an environment is created. 
First the channel state is sampled from users distributed uniformly 
across the system area as outlined in Section~\ref{user_distribution}. 
Next, the channel realization for each user is computed using the channel model described in Section~\ref{channel_model}. The output 
is the channel realizations in the time domain (i.e. the channel impulse response) which is converted to the frequency domain using an FFT and represented as a complex matrix with dimensions $N_\text{user} \times N_\text{subband} \times N_\text{rx} \times N_\text{tx}$.
The last step is to sample the buffer state for each user
by utilizing the simplistic traffic model outlined in Section~\ref{traffic_model}. 
This gives the buffer state (i.e. the number of bits in the downlink buffer) for each user. 

\vspace{-0.1cm}
\subsection{Monte Carlo Tree Search}\label{mcts}
MCTS is a family of algorithms tackling the complexity of searching in large tree-structured search problems by using Monte Carlo simulations. The results of such simulations are then used to iteratively focus the strategy towards more promising regions of the search space.

The search tree is progressively built starting from the root node through: \textit{selection} of one of the child nodes,  \textit{expansion} of a non-terminal child node (if such node has unvisited children), \textit{simulation} from the newly visited child node to a leaf node,  \textit{backpropagation} of the reward collected at the leaf node to update the statistics of the nodes in the path to the leaf.


The node statistics are updated as a function of at least two variables: the visit count and the reward value.
Such statistics are provided as input to the tree policy, i.e. the policy that determines the first two phases of selection and expansion.
A pivotal moment in the history of MCTS took place in 2006 when Upper Confidence Bound for Trees (UCT) \cite{UCT2006} was proposed.
In UCT the tree is explored by selecting the node that maximizes

\begin{align*}
\begin{split}
UCT(s, a) &= Q(s, a) + U(s, a)\\
&= \overline{R}(s, a) + 2 c_\text{puct} \sqrt{\frac{2 \ln N(s)}{N(s, a)}}, \\
\end{split}
\end{align*}
where $\overline{R}(s, a)$ is the average reward accumulated from the node with state~$s$ when choosing action~$a$ (a given arm),
$N(s, a)$ is the number of times action~$a$ has been selected and~$N(s)$ is the overall number of visits to the node so far.
UCT can thus be seen as the combination of a term $Q(s, a)$ that encourages exploitation of higher-reward choices and $U(s, a)$ that encourages exploration of less visited choices.
The constant $c_\text{puct}$ is used to adjust the balance of exploration and exploitation.

Even though the Monte Carlo simulations arise out of the necessity of traversing very large trees, many real world scenarios still remain out of reach due to the large combinatorial spaces and limited computational and time constraints.
These limitations have often been addressed by using domain-specific heuristics (on top of UCT-like policies) to bias action selection and further reduce the search space.
Such heuristics can either reduce the depth of the search tree (by replacing the first term $Q(s, a)$ with an estimation $\widetilde{Q}(s, a)$)
or the breadth of the search tree (by acting on the second term $U(s, a)$ excluding actions unlikely to lead to good outcomes).

The innovation introduced by \cite{Silver2017MasteringTG} consists of adopting deep learning techniques to learn the $Q$-value function and the policy $\pi$ from simulation and using the learned function inside the UCT formula.
This approach dramatically improves the capability of the algorithm to focus on promising regions (i.e., best-first) of the search space without the need for domain-specific heuristics.
This is possible because the deep neural network producing the functions $Q$ and $\pi$ does not learn independently for each node in the tree (as it is the case of the local statistics of UCT).
Instead, the same neural network is used across the tree allowing the experience of all nodes in the tree to be aggregated into a single function.

In \cite{Silver2017MasteringTG} the UCT formula was modified as follows to incorporate the output of the neural network:
\begin{align*}
\begin{split}
UCT(s,a) &= Q(s, a) + U(s,a)\\ 
&= Q(s, a) + c_\text{puct}P(s,a)\frac{\sqrt{\sum{_{a'}{N(s,a')}}}}{1+N(s,a)},
\end{split}
\end{align*}
where $Q(s, a)$ is the action-value prediction made by the neural network
and $P(s, a) = \pi(a|s)$ is the action prior probability given by the policy function.


\vspace{-0.1cm}
\subsection{Decision Evaluation}\label{decision_evaluation}
In this section we describe the method for calculating the reward, i.e. the performance metric to be optimized. We use the proportional fair metric to measure performance, but many other metrics, like sum throughput or minimum throughput could be used instead. The input is the environment state together with the final episode state (i.e. the scheduling decision). 

The first step is to calculate the precoder for each user and subband. Precoder calculation is based on the Signal-to-Leakage-and-Noise Ratio (SLNR)~\cite{slnr} and tries to balance the signal power to the wanted user with the suppression of interference from co-scheduled users. 
The precoder $\mtx{w}_{k,j}$ is calculated using 

\begin{equation}\label{precoder_eq}
\mtx{w}_{k,j} \propto \left(\mathbb{I}+\frac{1}{\sigma^2}\sum\limits_{i\neq k}\mtx{H}_{i,j}\mtx{H}_{i,j}^H\right)^{-1}\mtx{H}_{k,j},
\end{equation}
and normalized to unit power. Here, $\mtx{H}_{k,j}$ is the channel matrix for user~$k$ and subband~$j$, $A^H$ denotes the Hermitian transpose of A and $\sigma^2$ is the noise variance.
Only the transmitting users selected by the scheduling decision should be included in the sum in Eq.~\eqref{precoder_eq}. In this work we restrict it to two co-scheduled users. Hence there can be only one interfering user for a given subband.

Next, the SINR per user and subband is calculated as

\begin{equation}\label{sinr_eq}
SINR_{k,j}=\frac{\norm{\mtx{H}_{k,j}\mtx{w}_{k,j}}^2} {\sigma^2+\sum\limits_{i\neq k}\norm{\mtx{H}_{i,j}\mtx{w}_{i,j}}^2}.
\end{equation}
Based on the SINR the Transport Block Size (TBS) for each user is calculated. This is done using a simple link adaptation algorithm that tries to find the largest TBS for the given allocation that satisfies the Block Error Probability $\text{BLEP}_k < 0.1$. The set of available TBSes for each allocation size are taken from ~\cite[\mbox{Table 7.1.7.2.1-1}]{3gpp.36.213}.
The Mutual Information (MI) model~\cite{l2s} is used to calculate the BLEP for user~$k$. 

Next, the rate for user $k$ is calculated by scaling the TBS with the success probability and dividing by the transmission time as 
\begin{equation}\label{throughput_eq}
R_k = \frac{(1-\text{BLEP}_k)*\min(TBS_k,N_{bits,k})}{T_\text{TTI}}.
\end{equation}
Here, $\text{BLEP}_k$ is again calculated using the method in~\cite{l2s}, $N_{bits,k}$ is the number of bits in the buffer for user $k$ and $T_\text{TTI}$ is the time of a Transmit Time Interval (TTI).

To rate the quality of a scheduling decision based on a set of achieved rates $R_k$ a utility function is used. In this work we consider Mobile Broadband (MBB) traffic for which the Proportional Fair (PF) metric is suitable. If one instead would consider some type of delay sensitive traffic a utility metric based on delay would be more suitable.
The PF metric for user $k$ is defined as 
\begin{equation}\label{pf_metric_eq}
U_{PF,k} = \frac{R_k}{\overline{R}_k}.
\end{equation}
Here, $R_k$ and $\overline{R}_k$ are the instantaneous and average rate for user $k$ respectively. 


Finally, the reward is calculated as the total proportional fair metric scaled by a constant ($\beta$)

\begin{equation}\label{reward_eq}
r = \beta * \sum\limits_{k}{}{U_{PF,k}}.
\end{equation}
$\beta$ has been set to $0.1$ and is used to approximately normalize the rewards. 

The above steps are summarized in Alg.~\ref{decision_evaluation_alg}.

\begin{algorithm}[t!] \scriptsize
\caption{Decision Evaluation}\label{decision_evaluation_alg}
\begin{algorithmic}[t!]
\vspace{0.2cm}
\State \textbf{Input:} Environment state and Episode state
\vspace{0.1cm}
\State Calculate the:
\State \textbf{1.} Precoder for each user and subband (Eq.~\ref{precoder_eq})
\State \textbf{2.} SINR for each user and subband (Eq.~\ref{sinr_eq})
\State \textbf{3.} Transport Block Size ($TBS_k$) for each user
\State \textbf{4.} Block Error Probability ($BLEP_k$) for each user~\cite{l2s}
\State \textbf{5.} Throughput for each user (Eq.~\ref{throughput_eq})
\State \textbf{6.} PF metric for each user (Eq.~\ref{pf_metric_eq})
\State \textbf{7.} Reward as the total PF metric scaled by a constant (Eq.~\ref{reward_eq})
\end{algorithmic}
\end{algorithm}

\vspace{-0.1cm}
\subsection{Training with Reinforcement Learning}\label{training_RL}

The training procedure is outlined in Algorithm~\ref{training_alg}. It consists of two phases that are iterated, where the first phase generates \num{200} environment instances, solves them using \num{200} MCTS simulations and evaluates the resulting decision. For each environment instance the states ($\pmb{s}$), the search policy vectors ($\pmb{\pi}$) and the search values ($z$) are stored in a dataset. In a second phase this dataset is used to retrain the neural network using the loss function 

\begin{equation*}
l=(z-v)^2 + \pmb{\pi}^\intercal\log \textbf{p},
\end{equation*}
where $z$ is the search value, $v$ is the value prediction, $\pmb{\pi}$ is the search policy and $\textbf{p}$ is the policy prediction. We use the Adam optimizer with a learning rate of \num{1e-4} and train it over  \num{50} epochs in each iteration.

This is repeated for  \num{25} iterations after which a final evaluation is performed by solving \num{10000} new environment instances. Due to the stringent real-time requirements, no search is performed during the evaluation phase. Instead the action with the highest action probability is selected in a greedy manner. 

\begin{algorithm}[t!] \scriptsize
\caption{Training Procedure}\label{training_alg}
\begin{algorithmic}[t!]
\vspace{0.2cm}
\State \textbf{Input:} Untrained neural network 
\vspace{0.1cm}
\For {$iteration \in 1,\ldots,25$}
\State Dataset $\mathcal{D} \gets \{\}$
\For {$environment \in 1,\ldots,200$}
\State \textbf{1.} Sample environment (Section~\ref{state_sampling})
\State \textbf{2.} Solve environment using MCTS + Neural Net (Section~\ref{mcts})
\State \textbf{3.} Evaluate decision (Section~\ref{decision_evaluation}) 
\State \textbf{4.} Append state transitions $(\pmb{s},\pmb{\pi}, z)$ to $\mathcal{D}$ 
\EndFor
\State \textbf{5.} Train neural network with the generated data ($\mathcal{D}$)

\EndFor
\State \textbf{6.} Final evaluation on \num{10000} new environments
\end{algorithmic}
\end{algorithm}

\subsection{Policy and Value Prediction}\label{neural_network}

\begin{figure}[!ht]
\vspace{-0.2cm}
\centering
\resizebox{3.0in}{!}{\begin{tikzpicture}[font=\small,thick, every text node part/.style={align=center}]

\node[draw,
    rounded corners=3,
    minimum width=2.5cm,
    minimum height=0.75cm] (input) {State};

\node[draw,
    above=0.5cm of input,
    rounded corners=3,
    minimum width=2.5cm,
    minimum height=0.75cm] (dense1) {Dense};

\draw[-latex] (input) edge (dense1);

\node[draw,
    above=0.5cm of dense1,
    circle,
    minimum width=0.75cm] (add1) {\textbf{+}};

\draw[-latex] (dense1) edge (add1);

\node[draw,
    left=0.75cm of add1,
    rounded corners=3,
    minimum width=2.5cm,
    minimum height=0.75cm] (posenc) {2D Positional \\ Encoding};

\draw[-latex] (posenc) edge (add1);

\node[draw,
    above=1.25cm of add1,
    rounded corners=3,
    minimum width=2.5cm,
    minimum height=0.75cm] (mha) {Multi-Head \\ Attention};

\draw[-latex] (add1) edge node[xshift=-2, yshift=-4, pos=1, right]{$\scriptstyle K$} (mha);
\draw[-latex] 
  ([yshift=-5mm] mha.south) -- ++(5mm,0) -- node[xshift=-2, yshift=-4, pos=1, right]{$\scriptstyle Q$} ([xshift=5mm] mha.south); 
\draw[-latex] 
  ([yshift=-5mm] mha.south) -- ++(-5mm,0) -- node[xshift=-2, yshift=-4, pos=1, right]{$\scriptstyle V$} ([xshift=-5mm] mha.south);


\node[draw,
    above=0.5cm of mha,
    rounded corners=3,
    minimum width=2.5cm,
    minimum height=0.75cm] (addnorm1) {Add \& Norm};

\draw[-latex] (mha) edge (addnorm1);

\draw[-latex] 
  (add1.north) -- ++(0,+15pt) -- ++(-50pt,0pt) |- (addnorm1.west);

\node[draw,
    above=0.5cm of addnorm1,
    rounded corners=3,
    minimum width=2.5cm,
    minimum height=0.75cm] (feedfwd) {Pointwise \\ Feed-forward};

\draw[-latex] (addnorm1) edge (feedfwd);

\node[draw,
    above=0.5cm of feedfwd,
    rounded corners=3,
    minimum width=2.5cm,
    minimum height=0.75cm] (addnorm2) {Add \& Norm};

\draw[-latex] (feedfwd) edge (addnorm2);

\draw[-latex] 
  (addnorm1.north) -- ++(0,+5pt) -- ++(-50pt,0pt) |- (addnorm2.west); 
  
\draw [fill=black!10, opacity=0.25,rounded corners=3] ($(mha.south west)+(-19pt,-25pt)$) rectangle ($(addnorm2.north east)+(5pt,5pt)$);

\node[left=0.7cm of addnorm2] (nblocks) {x2};


\node[draw,
    right=1cm of dense1,
    rounded corners=3,
    minimum width=1.75cm,
    minimum height=0.75cm] (dense1p) {Dense};
    
\draw[-latex] 
  (addnorm2.north) -- ++(0,+10pt) -- ++(50pt,0pt) |- ($(dense1p.south)+(0,-15pt)$) -- (dense1p.south); 

\node[draw,
    above=0.5cm of dense1p,
    rounded corners=3,
    minimum width=1.75cm,
    minimum height=0.75cm] (dense2p) {Dense};

\draw[-latex] (dense1p) edge (dense2p);

\node[draw,
    above=0.5cm of dense2p,
    rounded corners=3,
    minimum width=1.75cm,
    minimum height=0.75cm] (dense3p) {Dense};

\draw[-latex] (dense2p) edge (dense3p);

\node[draw,
    above=0.5cm of dense3p,
    rounded corners=3,
    minimum width=1.75cm,
    minimum height=0.75cm] (softmax) {Softmax};

\draw[-latex] (dense3p) edge (softmax);

\draw[-latex] (softmax) edge node[pos=1.0, right]{$\mathbf{p}$} ([yshift=5mm] softmax.north);


\node[draw,
    right=0.5cm of dense1p,
    rounded corners=3,
    minimum width=1.75cm,
    minimum height=0.75cm] (dense1v) {Dense};
    
\draw[-latex] 
  ($(dense1p.south)+(0,-15pt)$) |- ($(dense1v.south)+(0,-15pt)$) -- (dense1v.south); 

\node[draw,
    above=0.5cm of dense1v,
    rounded corners=3,
    minimum width=1.75cm,
    minimum height=0.75cm] (dense2v) {Dense};

\draw[-latex] (dense1v) edge (dense2v);

\node[draw,
    above=0.5cm of dense2v,
    rounded corners=3,
    minimum width=1.75cm,
    minimum height=0.75cm] (dense3v) {Dense};

\draw[-latex] (dense2v) edge (dense3v);

\draw[-latex] (dense3v) edge node[pos=1.0, right]{$v$} ($(dense3v.north)+(0,1.8cm)$);

\end{tikzpicture}}
\caption{The Neural Network based on the encoder part of the transformer.}
\vspace{-0.2cm}
\label{fig:neural_network}
\end{figure}
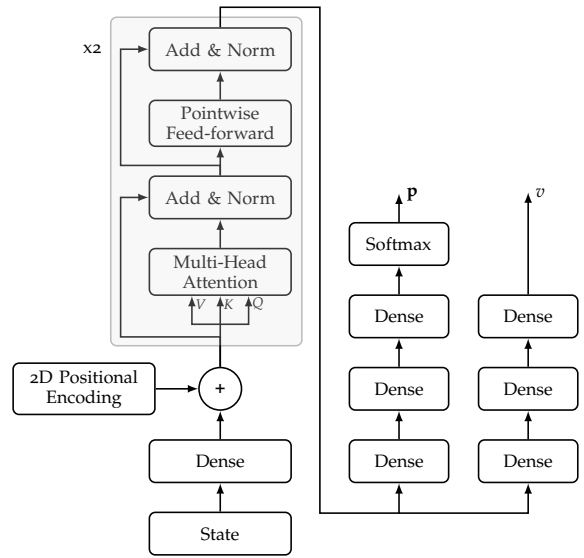

To make policy and value predictions, a multi-task neural network (depicted in Fig.~\ref{fig:neural_network}) is used. This neural network takes the pre-processed state and outputs a policy $\pmb{p}$ and a value v. The policy is given as action probabilities where the actions are the enumerated combinations of user allocations (as described in Section~\ref{mdp}). 
The neural network is used to steer the search conducted by MCTS to focus the search on the parts of the search tree where well performing solutions are likely to be found. 

The design is based on the encoder of a transformer~\cite{vaswani2017attention} with two encoder blocks and two attention heads each. We find that the self-attention mechanism utilized by the transformer is crucial for the policy prediction to be effective when the number of users becomes larger than two. 
Self attention efficiently captures dependencies between inputs regardless of the distance between them. This enables efficient reuse of local computations across permutations of users and resources. 

Both the policy and the value heads are MLPs with 2 hidden layers with 32 activations in each layer and Leaky ReLU activation function. The output layers for the policy head has $|A|$ activations whereas the value head is a scalar with \textit{softplus} activation function.

Since the self-attention is permutation equivariant, 2D positional encodings based on~\cite{wang2019translating2} are added to the input to the first self-attention block to maintain information that is encoded in the input ordering.

To be able to predict the policy and the value for a state, the model will have to learn how much co-scheduled users interfere with each other, given that the precoder for each user is selected according to some criterion (in our case the SLNR precoder). Hence, to make the prediction task easier the state needs to be represented in an appropriate way. 
For the complex channel matrices we look at pairs of users and represent the channel of each pair with three scalars, which are the magnitude of the dot product, the Hermitian angle ($\Theta_H$) and Kasner’s pseudo angle ($\varphi$), where the two latter ones are defined in~\cite{ComplexAngles}. 


\vspace{-0.15cm}
\section{Simulator Description}\label{simulator_description}

\vspace{-0.1cm}
\subsection{User Distribution}\label{user_distribution}
We define the system area as a circle sector with a radius of \num{500} m and a central angle $\alpha=65$\textdegree. The user height is set to \num{1.5} m for all users. The base station is positioned at the origin with the antenna array azimuth set to \num{0}\textdegree. The distance between the antenna and the user is clipped to always exceed \num{35} m which is the minimum distance for the UMa model described in section~\ref{channel_model}.

\vspace{-0.1cm}
\subsection{Channel Model}\label{channel_model}
We consider the channel model described in~\cite{PropagationModel}. This is a spatial channel model that covers carrier frequencies in the range \num{0.5}-\num{100} GHz. This model supports four scenarios, which are urban microcell street canyon (UMi), urban macrocell (UMa), indoor office, and rural macrocell (RMa). In this work we mainly consider the UMa scenario but also the UMi scenario for some evaluations. 

\vspace{-0.1cm}
\subsection{Traffic Model}\label{traffic_model}
For evaluations with finite buffer traffic we employ a simplistic traffic model where the amount of data (i.e. number of bits) in the downlink buffer for user $k$ is calculated as

\begin{equation*}
N_{bits,k} = 8 \cdot \left \lfloor b/8 \right \rfloor
\end{equation*}
where $b$ is uniformly distributed as \begin{equation*}
b\sim\mathcal{U}(N_\text{min},N_\text{max})
\end{equation*}

To make results easily comparable we also make sure that there is at least one user, selected randomly, for which the buffer is full. By doing this we can assume that an optimal solution should always schedule all subbands and we can therefore use rate-based metrics instead of e.g. spectral efficiency as performance metric.

\vspace{-0.1cm}
\subsection{Baseline Scheduler}\label{baseline_scheduler}
The baseline scheduler used in this work is based on the optimization method defined by \cite[Alg. 7.1]{ResourceAllocBook}, which is a strong heuristic algorithm with quadratic complexity. Some minor modifications are introduced to support scheduling of up to $M$  users per subband. 

Subbands are allocated to users in a way that maximizes the marginal utility, i.e. the gain in the utility $U_{PFTF,k}$ when an extra subband $i$ is allocated to user $k$, compared to the utility of user $k$ before the allocation of subband $i$. 

For optimization criterion we use the  Proportional Fair Time Frequency (PFTF) metric 
which can be written as

\begin{equation*}
    U_{PFTF,k} = 
\begin{cases}
    \frac{\sum\limits_{j\in\mathcal{I}_{sb,k} \cup \{i\} }{R_{k,j}^{}}}{\overline{R}_k+\sum\limits_{j\in\mathcal{I}_{sb,k}}{R_{k,j}^{}}},& \sum\limits_{j\in\mathcal{I}_{sb,k}}{R_{k,j}T_{TTI}^{}} \leq N_{bits,k} \\
    0,              & \text{otherwise}
\end{cases}
\end{equation*}
where $R_{k,j}$ is the rate for user $k$ in subband $j$ in the current TTI and $\overline{R}_k$ is the average rate of user $k$ over a time window. The instantaneous and average user rates are calculated based on Eq.~\eqref{throughput_eq}. 

\vspace{-0.1cm}
\subsection{Simulation Parameters}
In Table~\ref{enironment_parameters} the environment parameters can be found.

\begin{table}[t!]\footnotesize
\caption{Environment Parameters}\label{enironment_parameters}
\centering
\vspace{-0.20cm}
\begin{tabular}{p{3cm}p{3cm}} \toprule
\textbf{Parameter} & \textbf{Value} \\ \midrule
Channel Model & 3GPP Urban Macro \\
Carrier Frequency & 3.5 GHz\\
Bandwidth & 40 PRBs (8 MHz)\\
Subband Size & 4 PRBs \\
Deployment & Single Cell\\
User Speed & 0.1 m/s\\
Transmit Power & 0.8W/PRB \\
Noise Power & 112.5 dBm/PRB \\
Antenna Config & 2 Tx, 1 Rx \\
Transmit Time Interval & 1 ms \\ \bottomrule
\end{tabular}
\vspace{-0.4cm}
\end{table}

\vspace{-0.15cm}
\section{Simulation Results}\label{simulation_results}
\vspace{-0.1cm}
In this section we present results where we compare our approach to the PFTF baseline outlined in Section~\ref{baseline_scheduler}. Both scheduling strategies are evaluated on identical channel realizations and the normalized reward is calculated as the ratio between the reward for the trained scheduler and the reward for the baseline scheduler. During evaluation no MCTS simulations are performed but instead the highest ranking action from the neural net is selected.

\vspace{-0.1cm}
\subsection{Noisy Channel Estimates}
Due to non-ideal channel estimation there will be some residual errors in the channel estimates used for scheduling. We model these errors as complex Gaussian with a noise power $\sigma^2_{CE}$ calculated from the intended channel estimate SNR. 
\begin{equation*}
\sigma^2_{CE} = \norm{\mtx{H}_k}^2 \frac{1}{SNR_{CE}}
\end{equation*}
where $\mtx{H}_k$ is the true channel for user $k$.

The estimated channel is then calculated as 
\begin{equation*}
\hat{\mtx{H}}_k = \mtx{H}_k + n
\end{equation*}
with $n$ sampled from a complex Gaussian distribution:
\begin{equation*}
n {\sim} \mathcal{CN}(0,\sigma_{CE}^2)
\end{equation*}

\begin{figure}[!ht]
\vspace{-0.2cm}
\centering
\includegraphics[width=3.45in]{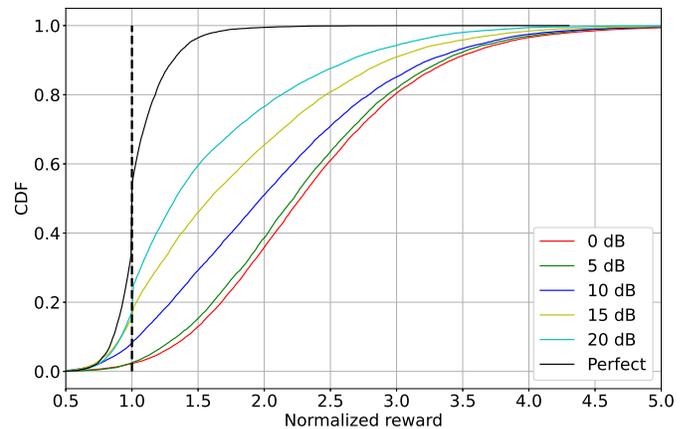}

\caption{Reward normalized to the baseline scheduler (dashed) 
in the presence of Noisy Channel Estimates. }
\vspace{-0.2cm}
\label{fig:noisy-csi-normalized-reward}
\end{figure}

Fig.~\ref{fig:noisy-csi-normalized-reward} shows a Cumulative Distribution Function (CDF) of the normalized reward when the channel estimates are corrupted by noise. Performance is shown for a set of channel estimate SNRs ($SNR_{CE}$) from \num{0} dB to perfect. The performance for the case with perfect channel estimates is close to the baseline performance and we interpret this as both solutions being fairly close to optimal. However, as the channel estimates become worse the relative performance of our approach increases and with \num{0} dB SNR the median performance gain is 125\%. This gain comes from the fact that our approach can decide to be more defensive in presence of uncertainty, for example not to co-schedule two users if their respective channels risk lining up unfavorably.

\vspace{-0.1cm}
\subsection{Channel Aging}
Due to the delay from the time when the channel is measured to when the channel estimate is used for scheduling, the channel information will be aged. If the user is moving fast this will cause more severe performance degradation. To model channel aging effects the channel is sampled twice with \num{10} ms delay between the samples. The first sample is used as channel estimate $\hat{\mtx{H}}_k$ and the second sample is the true channel $\mtx{H}_k$. These samples will be correlated and the correlation depends on the time between the samples (\num{10} ms), the user speed and the carrier frequency (\num{3.5} GHz). Also here, this gain comes from the fact that our approach can decide to be more defensive in presence of uncertainty.

\begin{figure}[!ht]
\vspace{-0.2cm}
\centering
\includegraphics[width=3.45in]{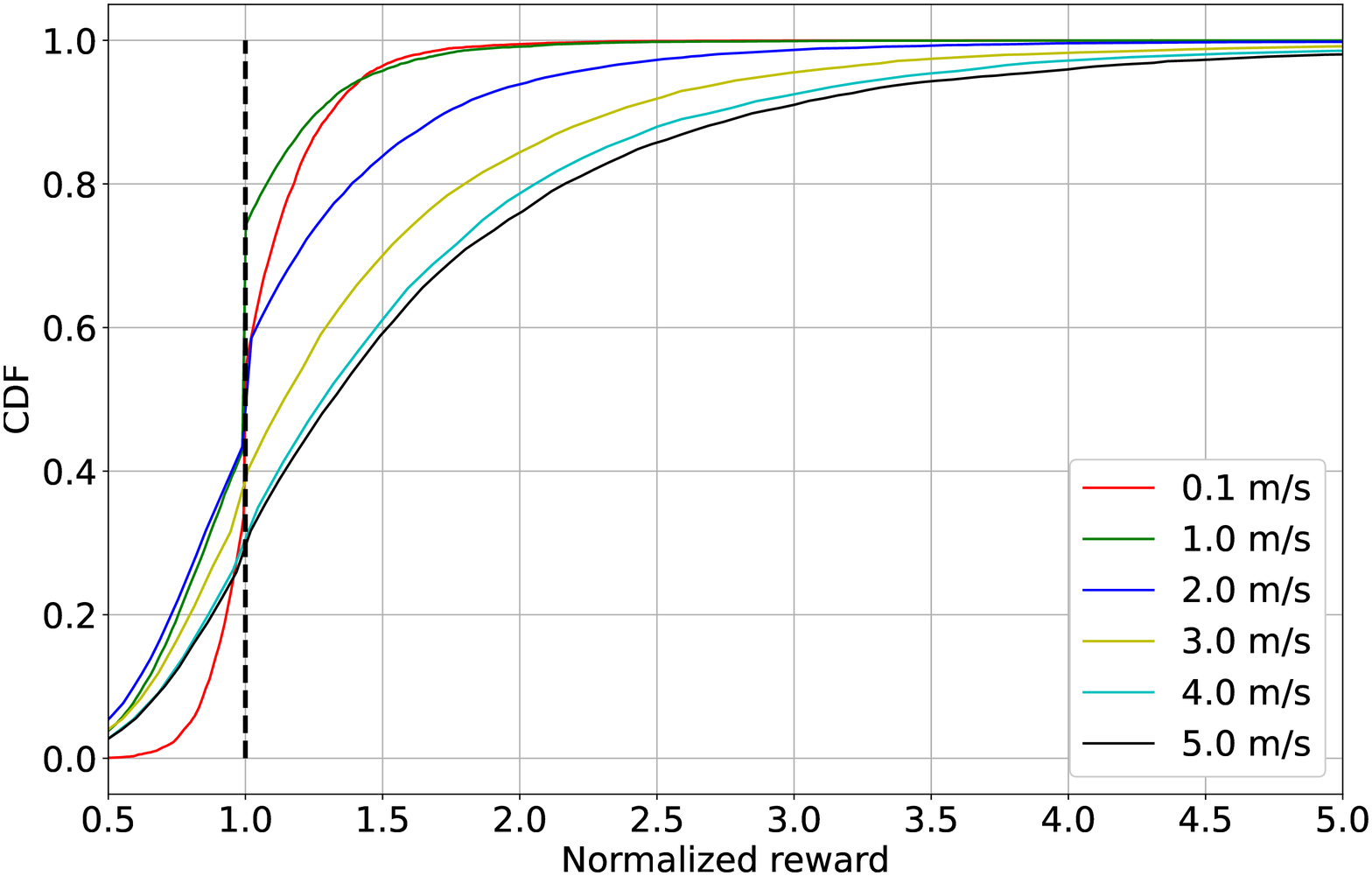}
\caption{Reward normalized to the baseline scheduler (dashed) 
in the presence of Channel Aging.}
\vspace{-0.2cm}
\label{fig:channel-aging-normalized-reward}
\end{figure}

Fig.~\ref{fig:channel-aging-normalized-reward} shows a CDF of the normalized reward when the channel estimate is an aged version of the true channel. The user speed is varied from \num{0.1} m/s to \num{5.0} m/s. The median performance gain is \num{30}\%  for a speed of \num{5} m/s.

\vspace{-0.1cm}
\subsection{Finite Buffer Traffic}
\begin{figure}[!ht]
\vspace{-0.2cm}
\centering
\includegraphics[width=3.45in]{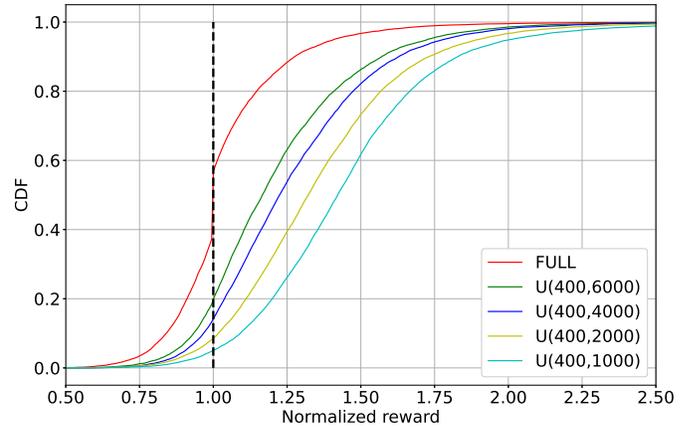}
\caption{Reward normalized to the baseline scheduler (dashed) 
for finite buffer traffic.}
\vspace{-0.2cm}
\label{fig:non-full-buffer-traffic-normalized-reward}
\end{figure}

Fig.~\ref{fig:non-full-buffer-traffic-normalized-reward} shows a CDF of the normalized reward in a set of finite buffer traffic scenarios. 
The legend indicates the distribution that buffer sizes are drawn from as described in Section~\ref{user_distribution} where e.g. "U(400,6000)" means that buffer sizes are uniformly distributed between 400 and 6000 bits.
It can be seen that with full buffer traffic the performance is close to the baseline. However, as users become more buffer limited gains become more pronounced. This can be explained by the fact that the baseline scheduler performs link adaptation for a first user without knowing if there will be interference from another user, whereas the learned scheduler use the global state when deciding on the user allocations. The median performance gain is 40\%  for a buffer state "U(400,1000)".

\vspace{-0.1cm}
\subsection{Generalization}
To make solutions like this useful in practice it is important that it can generalize to problem instances not experienced during training. 
To test this we look at the performance in terms of normalized reward when the model is trained on one set of channel models and evaluated on a different set. The model generalizes very well. The performance drop when training on UMa channels and evaluating on UMi channels is  less than 3\% as an example. 

\vspace{-0.15cm}
\section{Conclusions and Future Work}\label{conclusions}
\vspace{-0.1cm}
The results show that it is possible to train a scheduler with MCTS and Reinforcement Learning that outperforms a strong baseline scheduler in most scenarios. This is shown in particular when there is uncertainty in the channel state and when users have limited data in their buffers. The solution is also more future proof and easier to maintain due to the reduced need for domain knowledge.
In future work, inter-cell interference, scheduling across multiple TTIs and practical assumptions regarding number of users and cells should be considered.

\def\baselinestretch{0.87}
\bibliographystyle{IEEEtran}
\bibliography{references}
\vspace{-0.5cm}

\vspace{0.4cm}

\end{document}